\documentclass[preprint]{aastex}
\usepackage{graphicx,lscape}
\usepackage[usenames]{color}
\usepackage{float}

\newcommand{\kms}{\mbox{km\,s$^{-1}$}}
\newcommand{\lsun}{\mbox{${\cal L}_\odot$}\,}
\newcommand{\msun}{\mbox{${\mathrm M}_\odot$}}

\newcommand{\hi}{H\,{\sc i}}
\newcommand{\halpha}{H${\alpha}$}
\newcommand{\edit}{\color{black}}

\shorttitle{\hi\ asymmetries in the isolated galaxy CIG\,292}
\shortauthors{Portas et al.}

\begin{document}

\title{\hi\ asymmetries in the isolated galaxy CIG\,292}

\author{ Ant\'onio Portas\altaffilmark{1}, Tom C.\ Scott\altaffilmark{1}, Elias
Brinks\altaffilmark{2}, Albert Bosma\altaffilmark{3}, Lourdes Verdes--Montenegro\altaffilmark{1}, Volker Heesen\altaffilmark{2}, Daniel
Espada\altaffilmark{4}, Simon Verley\altaffilmark{5}, Jack Sulentic\altaffilmark{1}, Chandreyee Sengupta \altaffilmark{1,6}, E.\ Athanassoula\altaffilmark{3},  Min Yun\altaffilmark{7}}

\altaffiltext{1}{Instituto de Astrof\'{\i}sica de Andaluc\'{\i}a (CSIC), Glorieta de Astronom\'{\i}a s/n, 18008 Granada, Spain}
\altaffiltext{2}{Centre for Astrophysics, University of Hertfordshire, Hatfield, Herts AL10 9AB, United Kingdom}
\altaffiltext{3}{Laboratoire d'Astrophysique de Marseille (LAM), UMR6110, CNRS/Universit\'e de Provence/CNRS, Technop\^ole de Marseille
Etoile, 38 rue Fr\'ed\'eric Joliot Curie, 13388 Marseille CEDEX 13, France}
\altaffiltext{4}{National Astronomical Observatory of Japan (NAOJ),
2-21-1 Osawa, Mitaka, Tokyo 181-8588, Japan}
\altaffiltext{5}{Dept. de F\'isica Te\'orica y del Cosmos, Facultad de Ciencias, Universidad de Granada, Spain}
\altaffiltext{6}{Calar--Alto Observatory, Centro Astron\'omico Hispano Alem\'an, C/Jes\'us Durb\'an Rem\'on, 2-2 04004 Almeria, Spain}
\altaffiltext{7}{Department of Astronomy, University of Massachusetts, 710 North Pleasant Street, Amherst, MA 01003, USA}
\begin{abstract}

{\edit We present Expanded Very Large Array (EVLA) D-array
observations in the 21-cm line of neutral hydrogen (\hi) of
CIG 292,} an  isolated SA(s)b galaxy at a distance of
$\sim$ 24.3\,Mpc.  From previous \hi\ single dish observations the
galaxy was known to have a mildly asymmetric \hi\ profile (A$_{flux}$
=1.23 $\pm 0.3$). Our EVLA observations show there is  $\sim$ 12\%
more \hi\ projected South of the optical centre (approaching
velocities) than in the North (receding velocities), despite the
\hi\ extending $\sim$ 16\% further to the North than the South.
The \hi\ projected within the optical disk must have been perturbed
within the \hi\ relaxation time  ($\sim$10$^8$ yr ) which implies that
this can not have been caused by any of the three nearest companions,
as their distance ($\sim 0.5$\,Mpc) is too {\edit large. Neither \hi--rich companions nor tidal tails  
were found within our field of view and velocity range covered.} Our kinematical data
suggest that the inner part harbors an oval distortion whereas the
outer regions show signs of a modest warp. 
{\edit The mild asymmetry in the \hi\  global profile thus actually 
masks stronger asymmetries in the two--dimensional distributions of 
gas and star forming regions in this galaxy. Since the galaxy 
is isolated, this must predominantly be due to processes related to its 
formation and secular evolution.}

\end{abstract}


\section{Introduction}
\label{Introduction}

Galaxy evolution and the resulting $z=0$ properties of galaxies depend
on the environment they find themselves in. In order to quantify the
effect of environment (`nurture') on their morphology, structure,
nuclear activity, star formation properties, etc., one needs a well
defined sample of galaxies that are minimally perturbed by other
galaxies to provide a baseline of what are thought to be unperturbed
(or pure `nature') objects. The AMIGA (Analysis of the Interstellar
Medium of Isolated GAlaxies, \citet{ver05}; http://amiga.iaa.es)
project provides a statistically significant sample  of the most
isolated galaxies. It is a refinement of the original Catalogue  of
Isolated Galaxies \citep[CIG;][]{kara73}. Galaxies in the AMIGA sample
have not been involved in a major tidal interaction within the last
$\sim$3 Gyr \citep{ver05}. Quantification of the strength of tidal
interactions with neighboring minor companions and the local number
density is available for all galaxies in the AMIGA sample
\citep{verl07a,verl07b}. The AMIGA project has clearly established
that the most isolated galaxies have different physical properties,
even compared to what are generally considered field samples, in terms
of their morphology (asymmetry, concentration), L$_{FIR}$, the
radio\--FIR correlation, rate of AGNs or \hi\ asymmetry \cite[][and
  references therein]{ver10}.

Despite having much lower rates of \hi\  and optical asymmetry than
galaxies in denser environments \citep{durb08,espada11a}, some
galaxies in the AMIGA sample show appreciable asymmetry.
\cite{espada11a} studied the \hi\ profiles of a sample of 166 AMIGA
galaxies using an \hi\ asymmetry parameter A$_{flux}$ defined as the
ratio between the receding and approaching sides of the single dish
profile. They found the distribution 
of this parameter to be well described
 by the right half of a 
Gaussian distribution, with only 2\% of the sample having an asymmetry
parameter in excess of 3$\sigma$ (A$_{flux}$ $>$ 1.39, meaning a 39\%
excess of flux in one half of the spectrum). They did not find any
correlation between the \hi\ asymmetry parameter and minor companions,
measured both as the tidal force (one\--on\--one interactions) and  in
terms of the number density of neighboring galaxies. In contrast,
field galaxy samples deviate from a Gaussian distribution and have
higher  (10\--20\%) rates of asymmetry \citep{espada11a}.
 
To investigate the causes of asymmetry in the absence of major
interactions we are carrying out resolved \hi\ studies of a sample of
AMIGA galaxies with the EVLA, covering a representative range of
\hi\ asymmetry parameters (A$_{flux}$ = 1.05 to 1.25). For example,
  \cite{espada05} presented \hi\ VLA D--configuration results of the asymmetric
  galaxy CIG\,96, with \cite{espada11b} giving results of further
  C--array observations of this galaxy. 
{\edit CIG\,96 has a large and asymmetric  \hi\ envelope
(R$_\mathrm{HI}$/R$_{25}$= 3.5),
which partly coincides with faint UV emission with a more regular 
distribution. The kinematics of this  \hi\ envelope shows an area of 
non-circular motions, which could not be attributed to a major interaction.}

   Here we present the results of
EVLA \hi\ mapping of CIG\,292 (NGC\,2712) which has an A$_{flux} =
1.23 \pm 0.30$, and can be considered as intermediate between 
symmetric and asymmetric objects such as CIG\,96.

 We ascertained once more that CIG\,292 is an isolated galaxy. Despite it being listed as part of a group \citep{tully08,maka11},
only 3 potential companions have similar recession velocities to
CIG\,292 but their projected distances are large (58\arcmin,
62\arcmin, and 67\arcmin,  thus all of order 0.5 Mpc). CIG\,292 turns
out to be isolated according to the criteria used by \citet{kara73},
with local number density and tidal strength characteristics falling
within the limits of the {\em bona fide} AMIGA sample \citep[$\eta_k =
  1.668$, Q$_{Kar}$ = -3.106;][]{verl07b}. No \hi\ rich (dwarf)
companions were detected within the 32\arcmin\ region mapped by the
EVLA primary beam either. Any current tidal effect is therefore
negligible and any past interaction would have been at least 3\,Gyr
ago.

The neutral hydrogen component of CIG\,292 has previously been
observed with the Westerbork Synthesis Radio Telescope (WSRT) by
\cite{krum82},  and with single dish telescopes by \cite{huc85} and
\cite{spr05}. CIG\,292 is particularly interesting since  the
    \hi\ disk is significantly more extended than the optical one (see
    Sect.~\ref{HI results}), and faint optical emission is found at large radius as
    well \citep{koop06}.

Section \ref{observations} gives details of the observations, with the
results in Section \ref{results} and discussion and concluding remarks
in Section \ref{discussion}. We calculate the distance to the galaxy,
based on the observed recession velocity and corrected for
Virgo--flow, to be   23.4\,Mpc which implies an angular scale where
1\arcmin\ corresponds to $\sim 7$\,kpc.

\section{\hi\ observations}
\label{observations}

\begin{table*}[htdp]

\centering
{\footnotesize
\caption{General properties and observational setup for CIG\,292 }
\begin{tabular}{ll}
\hline
Parameter &Value \\
\hline
Object  &  CIG\,292                                        \\
$\alpha$ (J2000)\tablenotemark{a} & $8^h 59^m 30.53^s$\\
$\delta$ (J2000)\tablenotemark{a} & 44\arcdeg 54\arcmin 51.5\arcsec\\ 
Morphological type\tablenotemark{b} & SA(s)b\\
Optical inclination\tablenotemark{b} & 58\degr\\
log L$_B$\tablenotemark{c} &  9.77 log \lsun \\ 
Instrument & EVLA                                                 \\
Observing date & 25$^\mathrm{th}$ April 2010                                  \\
Configuration & D--array                                      \\
Project ID& AE175                                            \\
Primary calibrator  &  3C147                                  \\
Secondary calibrator & J0834+5534                               \\
Central velocity &1842\,\kms                                  \\
FWHM of primary beam & 32\arcmin                                 \\
Total bandwidth & 4.0\,MHz                                      \\
Number of channels & 256                                        \\
Channel spacing &15.625\,kHz (3.32 \kms)          \\
FWHM of synthesized beam & 46\arcsec $\times$ 42\arcsec       \\
Time on source  & 3.2\,hr                                  \\
rms noise &1.25\,mJy\,beam$^{-1}$ (0.4\,K T$_\mathrm{B}$)      \\
\hline
\tablenotetext{a}{\citet{leon03}}
\tablenotetext{b}{\citet{durb08}}
\tablenotetext{c}{\citet{espada11c}}
\end{tabular}
\label{setup}
}
\end{table*}

Observations  of CIG\,292 were obtained using 26 antennas of
the NRAO\footnote{The National Radio Astronomy Observatory is a
  facility of the National Science Foundation, operated under
  cooperative agreement by Associated Universities, Inc.} Expanded
Very Large Array (EVLA) in D\--configuration on 25$^\mathrm{th}$
April, 2010 (Project ID: AE175). The primary calibrator 3C147 (assumed flux density of
22.5\,Jy) was observed at the beginning of the run; it was also 
 used as bandpass calibrator. Interspersed  with the source, the
secondary calibrator J0834+5534 (with a derived  flux  density of
$8.80\pm0.017$\,Jy) was observed. 
In total 3.2 hours were spent on source with  both polarizations 
recorded. CIG\,292 was observed across a spectral  bandwidth of 
4\,MHz (corresponding to $\sim 800$\,\kms), centered at a frequency 
of 1411.65 MHz, at a frequency resolution of 15.625\,kHz (3.32\,\kms).
{\edit In Table \ref{setup}, we
provide some general properties of the galaxy, as well as
the principal parameters of the observational setup.}

The data {\edit were} correlated using the new EVLA WIDAR
correlator, and the data reduction was performed using CASA (Common
Astronomy Software Applications)  version 3.02. For guidance we
followed the spectral line tutorials available online at the CASA
website\footnote{http://casaguides.nrao.edu/}. The data reduction was
carried out on one of the nodes of the cluster of the Centre for
Astrophysics Research at the University of Hertfordshire. 

The EVLA observations were carried out at night, avoiding any
possible issues related to Solar RFI. Antenna position corrections were applied
to antennas 12 and 22. We edited (flagged) the data based on the
average of the inner 150 channels (from the 256 available). We also
decided to omit all data from antennas  9, 14, 17, and 23 at this
stage as their L--band receivers had not been upgraded and their
visibilities presented values one order of magnitude lower than
average. 

Calibration of the spectral line data was performed by first solving
for phase corrections to the 3C147 scan followed by deriving the
bandpass complex gains on this calibrator. Amplitude and phase
solutions for the flux calibrator and phase calibrator (J0834+5534)
were derived on the average of channels  50 to 200 (i.e., on a
quasi--continuum data set). After first inspection, a further
iteration of flagging and calibration was applied, including flagging
of bad visibilities related to the target. Continuum subtraction was
performed using 10 channels on either side of the velocity range
covered by the target that were found to be free from line emission.

Imaging of CIG\,292 was performed using CASA's standard cleaning
algorithm. We  produced a cube using the Briggs robust weighting
scheme \citep{bri95}  where we found robust=0.5 to be a good
compromise between spatial resolution (which results in a beam size of
46\arcsec $\times$ 42\arcsec) and sensitivity. The cube was cleaned
down to a flux threshold of $3\sigma$ (3.75 mJy\,beam$^{-1}$).

The cube was subsequently transferred into AIPS (version 31DEC11) in
order to restore the velocity information to the header since
observations were taken at a fixed frequency. We fixed  the centre of
the band at a velocity of 1842 \kms. We separated genuine emission
from noise following the conditional blanking method employed by
\citet{wal08}. A data cube convolved to 75\arcsec\ resolution was used
to mask  regions of emission that were above $2\sigma$ rms noise over
three consecutive velocity channels. As a final step, the zero and
first order moment maps (total surface brightness and velocity field)
were created using AIPS task XMOM. Both the data cube and the moment 0
map were corrected for primary beam attenuation.

\section{EVLA \hi\ Results}
\label{results} 

\subsection{\hi\ integrated map and spectra}
\label{HI results}

\begin{figure}[H]
\centering
\includegraphics[width= 17cm,trim=10 10 10 10, clip ]{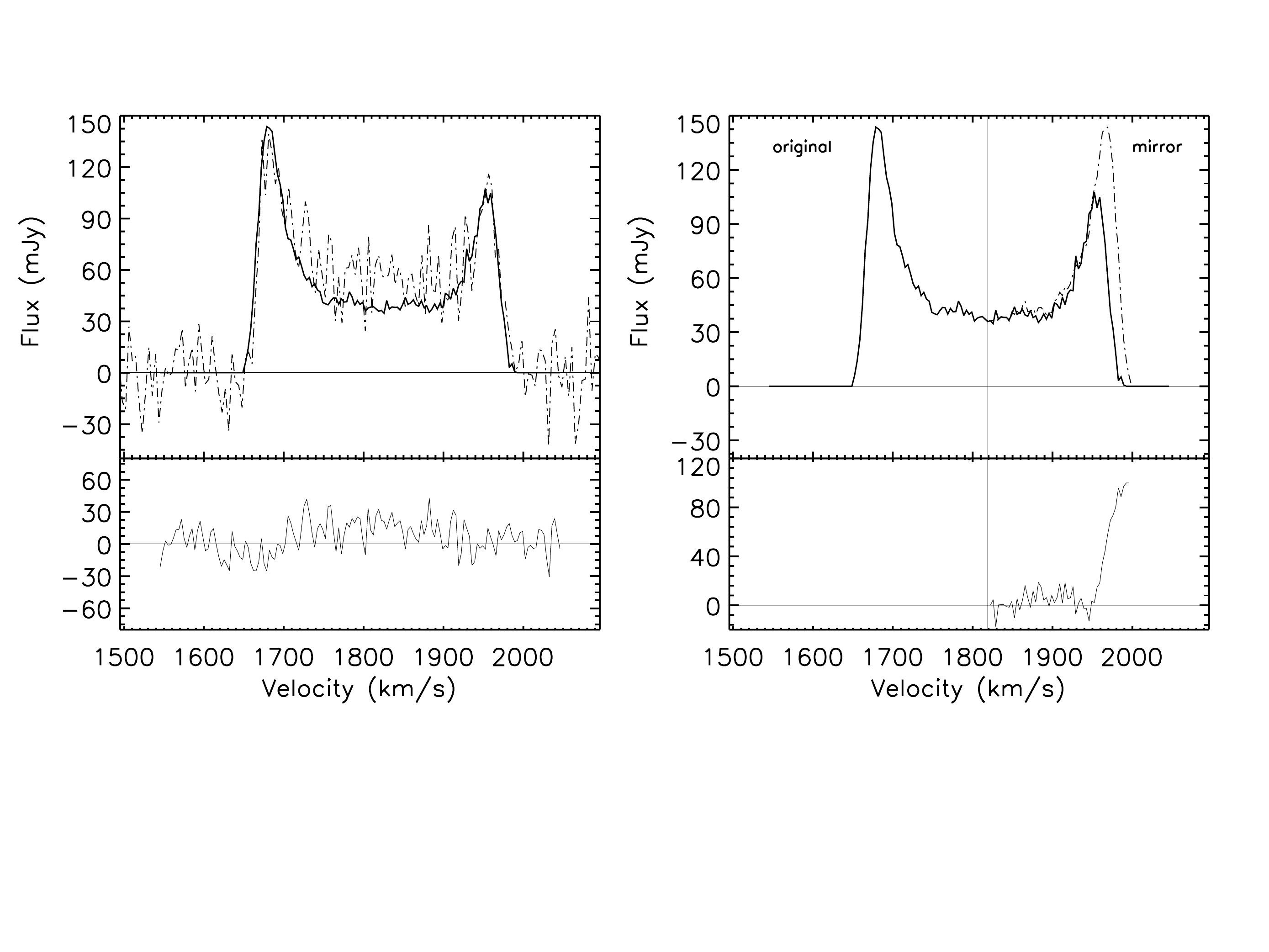}

\caption{Left panel: {\edit Integrated} \hi\ spectrum of CIG\,292 derived from the EVLA
  data (solid line). Emission is detected in the velocity range
  1650\,\kms\ to 1950\,\kms. The dashed dotted line represents the 43--m Greenbank single dish \hi\ profile retrieved from \citet{spr05}. Right panel: Integrated EVLA \hi\ spectrum of CIG\,292 {\edit where the approaching portion of the spectrum has been flipped around  V = 1818\,\kms\ (vertical line) and over--plotted (dashed--dotted line) for comparison with the receding portion of the spectrum}. The lower panels show the difference between each pair of spectra.} 
\label{figure1}
\end{figure}

We present in  Fig.\ref{figure1} the EVLA \hi\ integrated spectrum for
CIG\,292 and that of the  single dish profile obtained by \cite{spr05}
using the 43\--m Greenbank (formerly known as 140--ft) radio telescope
with a beam of
  20\arcmin. We also present in the same figure  the difference between
  both spectra.  We measured an integrated EVLA flux of 19.0 Jy \kms\,
  which translates to an \hi\ mass of 2.5 $ \times
  10^9$\,\msun\ assuming optically thin emission. This value is in
  good agreement with previous interferometric observations of this
  object performed with the WSRT
  \citep{krum82} of 20.1 Jy \kms\,(2.6 $\times 10^9$\,\msun) and the
  flux retrieved by \cite{spr05} obtained with the 43--m telescope of
  22.26 Jy \kms, the interferometers possibly missing a small fraction
  of low--level, extended emission. In that context it is interesting
  to note where the main differences between Greenbank and EVLA
  spectra originate (see Fig.~\ref{figure1}). We recover most of the flux
  in  the regions of the peaks where the emission in individual
  channels is most compact. It is in the central velocity channels
  (between 1700\,\kms \,and 1900\,\kms) of the galaxy where the
  \hi\ is  possibly more extended since on average 
  we underestimate the  single dish flux by some 20\%.

We also present in Fig.~\ref{figure1}, a folded profile determined
down to 20\% of the peak level, by flipping the emission around a 
velocity of 1818\,\kms. This shows that most of the asymmetry is 
coming from the highest velocities in the profile,  in the
1950\,\kms\ to 2000\,\kms\ range,  corresponding to the 
northern side of the galaxy.

\begin{figure}[H]
\centering

\includegraphics[width= 15 cm,angle=-90, trim=0 100 0 100, clip]{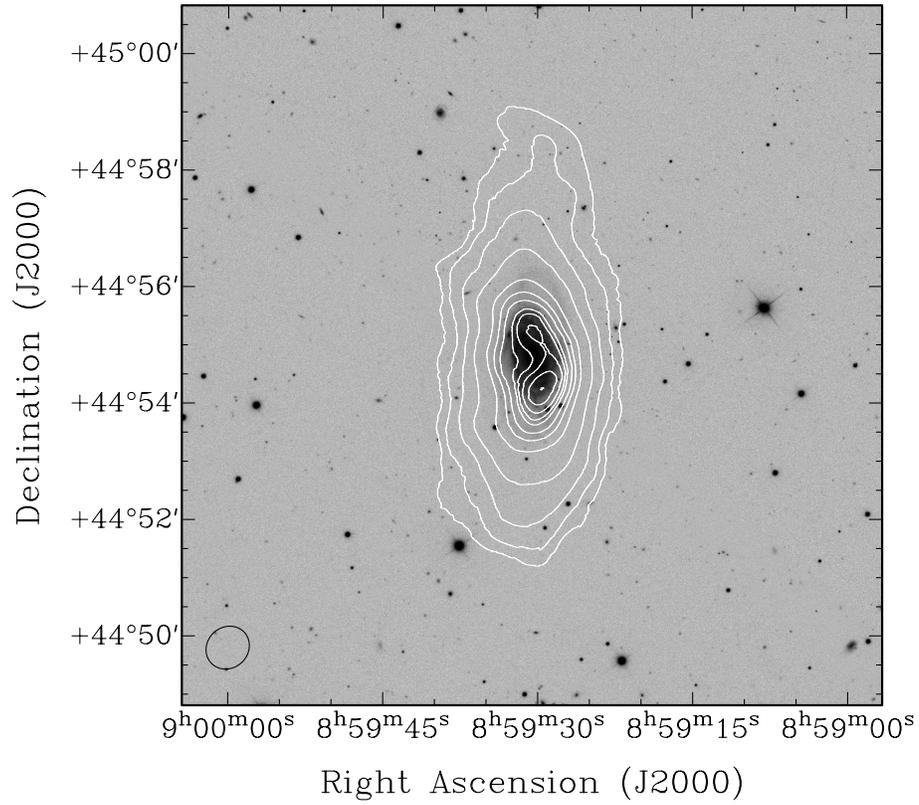}
\caption{\hi\ contours of CIG\,292 on an SDSS \textit{r--}band image. Contours are drawn at 0.5, 1.0, 1.2, 3.7, 6.2, 7.6, 8.5, 9.5, 10.0, 10.6,
 11.4, and 12.1 $\times 10^{20}$\,at\,cm$^{-2}$. The beam size is shown in the bottom left corner of the image. }
\label{figure2}
\end{figure}

In Fig.~\ref{figure2} we show the \hi\ column density map of CIG\,292
as contours overlaid on the SDSS $r-$band image. The galaxy has an
extended \hi\ disk with a diameter of $\sim$ 46\,kpc (6.5\arcmin)
measured at a {\edit column} density of $5 \times
10^{19}\,$at\,cm$^{-2}$. The new EVLA data are vastly superior to the
\citet{krum82} maps, the total \hi\ surface brightness map going
deeper by an order of magnitude, and the velocity resolution being
better by about a factor of 5. The specific benefits of the EVLA over
the VLA for this kind of work is, of course, the availability of the
vastly superior WIDAR correlator which offers a wide instantaneous
velocity range which opens up the possibility to search for companions
within a large volume around a target while maintaining high velocity
resolution to study any object thus detected.

Down to the noise level of $3.5 \times 10^{19}\,$at\,cm$^{-2}$ (a
$3\sigma$ detection across 5 channels or 16.5\,\kms\ which corresponds
to the typical line width of \hi\ at a velocity dispersion of
7\,\kms), the galaxy measures 3.6\arcmin\ (25\,kpc) to the South  and
4.3\arcmin\ (30\,kpc) to the North, making it lopsided. The extended
nature of the \hi\ structure in the North can be traced across 10
consecutive individual velocity channels ($\sim$ 32 \kms),  indicating
that it is part of the galaxy. The
R$_{25}$--to--R$_\mathrm{HI}$ ratio is $\sim 2.2$ assuming an R$_{25}$
of 1.5\arcmin\ (LEDA) and our R$_\mathrm{HI}$ of 3.25\arcmin\ measured
at $5 \times 10^{19}\,$at\,cm$^{-2}$. This falls within the typical range of
optical to \hi\ extent from the study by \citet{bro97} of
$1.7\pm0.5$. At the current \hi\ resolution, no details such as
  spiral arms or evidence for a bar can be discerned  in our
  \hi\ data.  {\edit Because
    up to 20\% of the total flux is missing in some of the
    channel maps due to the lack of short--spacings, which  mainly
    affects the extended structure, our  R$_\mathrm{HI}$ and
    R$_\mathrm{HI}$--to--R$_{25}$ ratio are lower limits. For the same
    reason the \hi\ mass to luminosity ratio presented in the
    following section is a lower limit.}

The inner disc of the galaxy is shown in more detail in
Fig.~\ref{figure3}, where we zoom in on the central region of the
galaxy, overlaying \hi\ contours on an SDSS $u, g$ and $r$ composite
image. Fig.~\ref{figure3} shows that the column density peak is offset
by 40\arcsec\ (4.6\,kpc) in projection to the SW of the optical centre
(marked with a star in the figure), with the higher density contours
forming an arc--like structure along the western side of the galaxy.

\subsection{Kinematics}

We present in  Fig.~\ref{figure4} the velocity field of CIG\,292. The
disk shows regular rotation throughout most of the disc. The
isovelocity contours at the outskirts (beyond 2.5\arcmin) are however
twisted, symptomatic of a warped disk. Initially the twist is point
symmetric and clockwise, but in the far North (beyond 3.6\arcmin\ or
25\,kpc) the position angle twists back, counter--clockwise.

A Position--Velocity (PV) diagram along the major axis of the 
galaxy (PA$=-5$\degr) is shown in Fig. \ref{figure4} (bottom panel). 
We trace the southern side  here out to $\sim$ 3.6\arcmin\ and the
northern side to $\sim$ 4.3\arcmin. The  (PV)--diagram shows steeply
rising rotation, unresolved by the beam, within the inner part (up to
$\sim$ 1\arcmin\ in diameter) followed by flat rotation out to the
last measured point on either side. \cite{marq04} using optical
long--slit spectroscopy confirm the steep rise in the gas component,
the peak velocity being reached within 10\arcsec\ (1.2\,kpc).

Assuming circular rotation and the inclination of 58\degr\ from LEDA,
we can use our data to find a dynamical mass for CIG\,292. At the last
point of the rotation curve at 30.1\,kpc (measured at a column density
of $3.5 \times 10^{19}\,$at\,cm$^{-2}$) we measure an indicative
rotation velocity of 120\,\kms, or an intrinsic velocity of
140\,\kms. Using M$_\mathrm{dyn} = 0.76 R V^2 /\mathrm{G}$ gives a
dynamical mass of $1.1 \times 10^{11}$\,\msun. This implies an
M/L$_\mathrm{B} = 18.9$. We find an M$_\mathrm{H}$/L$_\mathrm{B} =
0.4$.  This makes it underluminous for its Hubble type
\citep{rob94}. Further data at higher angular resolution are needed to
provide a detailed mass model.

\section{Discussion and concluding remarks}
\label{discussion}

The \hi\ results show that CIG\,292   has an asymmetric 
\hi\ distribution, extending farther North
than South,  with a twist in the outer
parts suggestive of a slightly warped atomic gas disk. Even
so, the \hi\ extension is at the side where there is the {\it least}
\hi\ in the integrated profile.  Thus, the asymmetry in the \hi\
global profile is partly masking a stronger asymmetry in the two-dimensional distribution,
{\edit as already suggested by \citet{espada11a}.}

A closer look at the available optical images in NED shows 
a system of faint optical arms \citep{koop06}. 
{\edit Using ellipse fitting, the position angle at those radii is 
about -3 degrees, in reasonable agreement with the 
PA of the major axis indicated by the kinematics, which is about -6
degrees between 2.5 and 4.0 arcmin radius. The more central parts
of the galaxy have a position angle of about +5 degrees, which 
indicates that the galaxy there has an oval shape, consistent 
with the fact that the kinematical major and minor axes at those
radii are not perpendicular
\citep{bosma78}. Moreover, this oval is seen as an
enhancement above the underlying exponential disk (a ``lens''
component) 
in the radial luminosity profile presented in \citet{erwin08}.
Such a feature is rather common in early type spiral galaxies
\citep[e.g.,][]{kor79}.}

\begin{figure}[H]
\centering
\includegraphics[width= 17 cm, trim= 20 100 20 100, clip ]{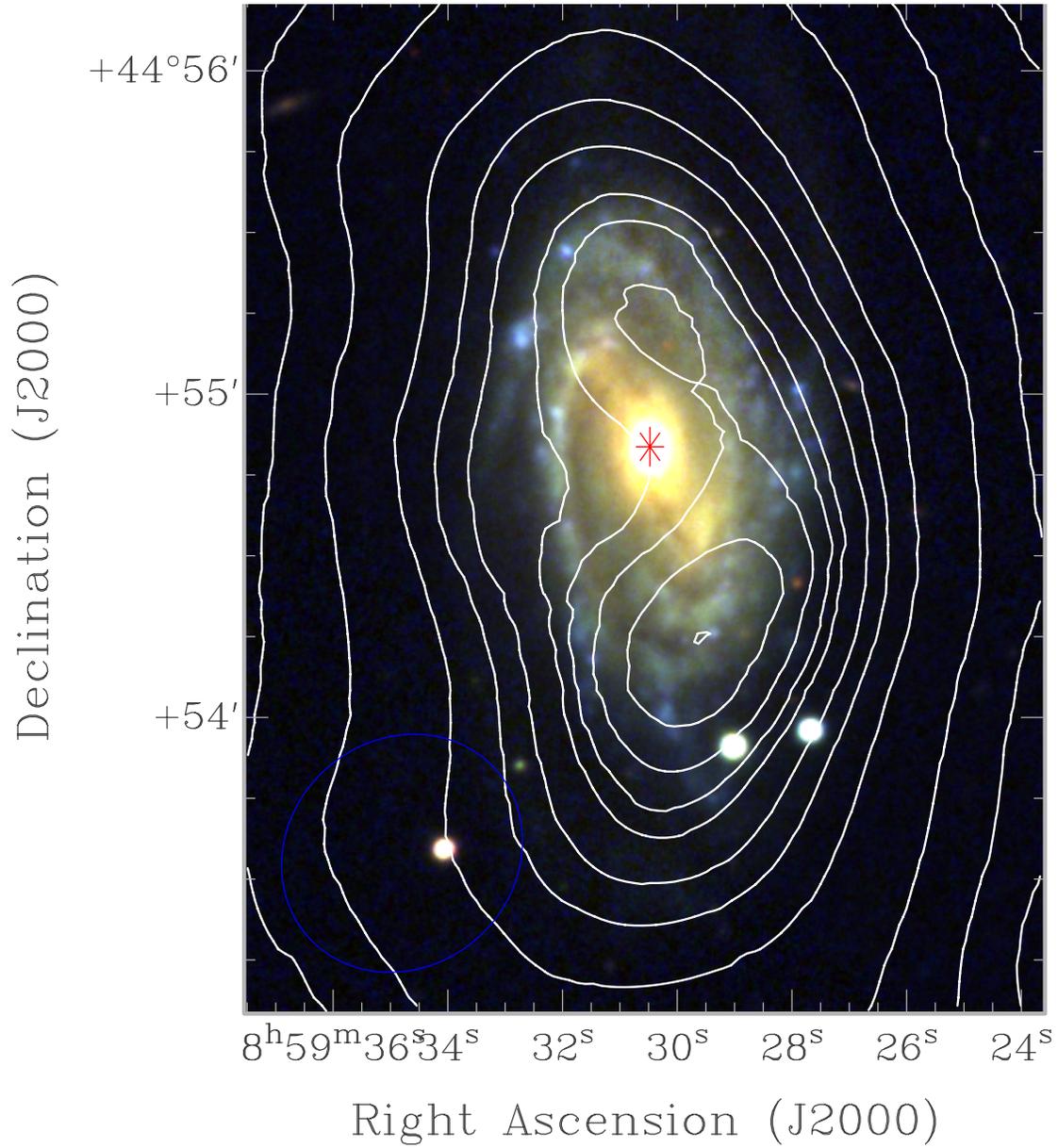}
\caption{Optical SDSS \textit{urg} composite of CIG\,292 with overlaid \hi\ contours. The  contours are presented as white solid lines  and correspond to  levels of 1.2, 3.7, 6.2, 7.6, 8.5, 9.5, 10.0, 10.6, 11.4, and 12.1 $\times 10^{20}$\,at\,cm$^{-2}$. The optical  center of the galaxy is indicated with a red star and the beam is shown in the bottom left of the image. }
\label{figure3}
\end{figure}

\begin{figure}[H]
\centering
\includegraphics[width= 16 cm, angle=-90, trim= 0 0 0 0, clip]{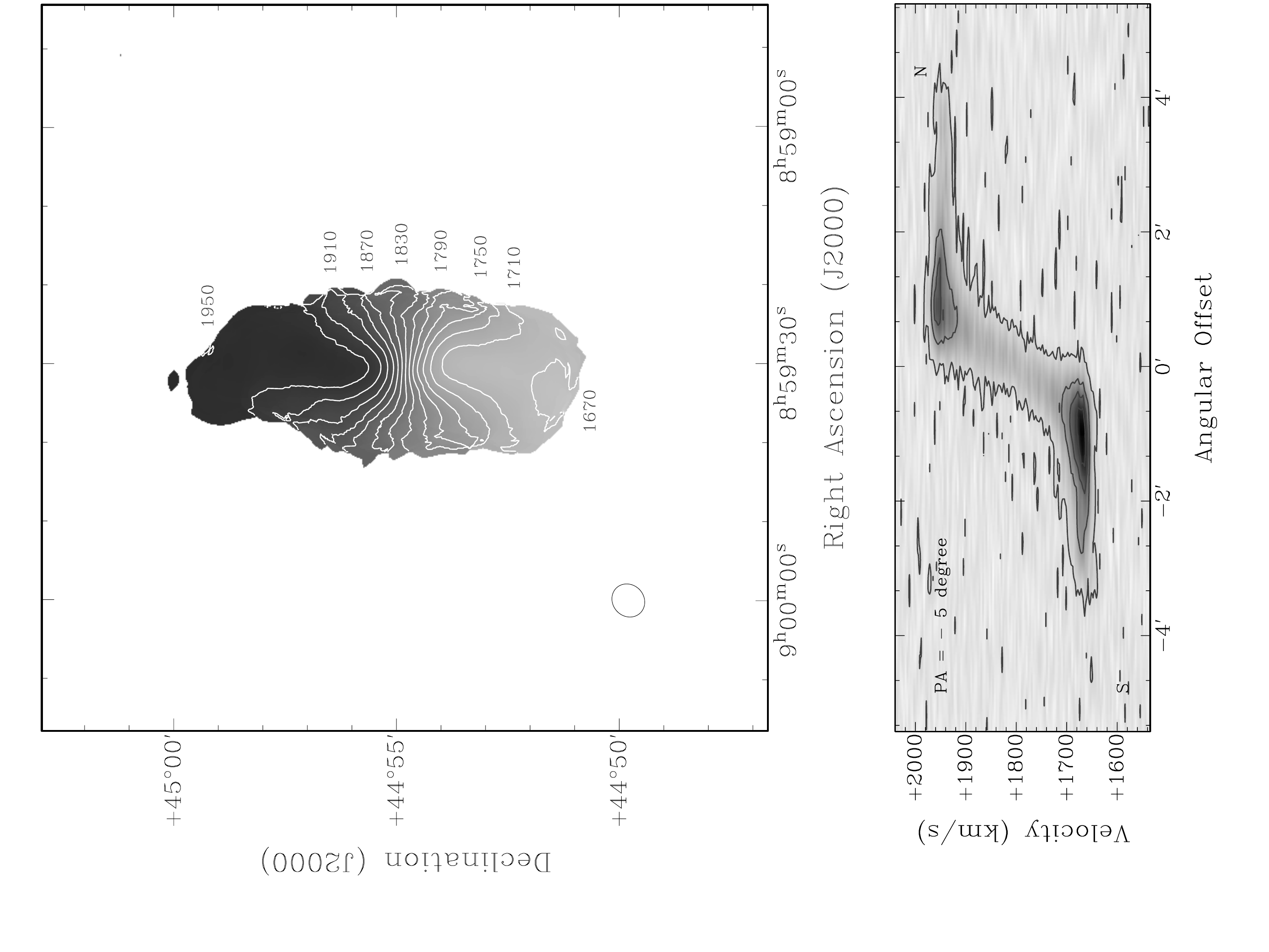}
\caption{Top panel: Velocity field of CIG\,292 with velocity contours superimposed. Velocity contours range from 1670\,\kms\  (southern half) to 1950\,\kms\ (northern half) in steps of 20\,\kms. We only label every other contour.  Bottom panel: Position--Velocity diagram along the major axis taken at a PA of -5\degr. Contours are at a level of 0.25, 0.4, 1.4, 2.4, 3.4 and 4.4\,mJy\,beam$^{-1}$.}
\label{figure4}
\end{figure}

Inspection of available {\em GALEX} images, \halpha\ images and the 8\,$\micron$ 
image from the {\em Spitzer} archive shows that this bright 
inner region is copiously forming stars, and has asymmetries in
the distribution of star forming regions. This coincides with the
asymmetry in the inner \hi\ distribution as noted in Fig.~\ref{figure3}.
These asymmetric distributions within the optical radius must have 
been perturbed within the relaxation time scale for the high density 
gas, i.e.,  of the order of 10$^8$ yr \citep{bosel94}. This short 
relaxation time scale seems to preclude an interaction with any of 
the three nearest companions as the source of the high--density 
\hi\ asymmetry.

We can exclude  as well an encounter with a companion with 
a large pericenter distance since this would necessitate a massive 
companion which is simply not there.  Alternatively the large scale 
asymmetry may be due to an interaction with a small companion which 
in the meantime has merged, or to ongoing cold accretion \citep[e.g.,][]{bourn05}.
 
In conclusion, our new EVLA data on this galaxy have revealed that
the mild asymmetry in the \hi\ global profile actually hides stronger 
asymmetries in the two-dimensional distributions of gas and of the star 
forming regions in this galaxy.  Our data indicate that the
central bright parts of the optical disk form an oval distortion, with
a position angle that is different from that of the faint optical disk and the
bulk of the \hi\ disk. In the northern part there is an extension, which is not 
necessarily in the same plane as the outer disk.

{\edit As for the causes of such asymmetries in isolated galaxies, the
oval distortion comes naturally from a bar instability in the disk
\citep[e.g.,][]{atha10}. The asymmetry in the neutral
gas distribution could be related to ongoing accretion of the
neutral gas from the cosmic web at angles forcing the outer \hi\ disk
to be warped w.r.t.\ the optical disk \citep[e.g.,][]{jiang99}.
Alternatively, it could be due to an m=1 instability. Data at higher angular 
resolution will be necessary to study the kinematics of the inner bright regions, 
in order to discern the dynamical causes of the asymmetries.}

\acknowledgements

This work has been supported by the research projects
AYA2008-06181-C02 from the Spanish Ministerio de Ciencia e Innovaci\'on,
the Junta de Andaluc\'{\i}a (Spain) grants P08--FQM--4205, FQM--0108,
TIC--114 and FP7--ICT--2009--6.

The Nasa Extragalactic Database, NED, is operated by the Jet
Propulsion Laboratory, California Institute of Technology, 
under contract with the National Aeronautics and Space Administration.

We acknowledge the usage of the HyperLeda database (http://leda.univ-lyon1.fr). 

Funding for the SDSS and SDSS-II has been provided by the Alfred
P. Sloan Foundation, the Participating Institutions, the National
Science Foundation, the U.S. Department of Energy, the National
Aeronautics and Space Administration, the Japanese Monbukagakusho, the
Max Planck Society, and the Higher Education Funding Council for
England. The SDSS Web Site is http://www.sdss.org/. The SDSS is
managed by the Astrophysical Research Consortium for the Participating
Institutions. The Participating Institutions are the American Museum
of Natural History, Astrophysical Institute Potsdam, University of
Basel, University of Cambridge, Case Western Reserve University,
University of Chicago, Drexel University, Fermilab, the Institute for
Advanced Study, the Japan Participation Group, Johns Hopkins
University, the Joint Institute for Nuclear Astrophysics, the Kavli
Institute for Particle Astrophysics and Cosmology, the Korean
Scientist Group, the Chinese Academy of Sciences (LAMOST), Los Alamos
National Laboratory, the Max-Planck-Institute for Astronomy (MPIA),
the Max-Planck-Institute for Astrophysics (MPA), New Mexico State
University, Ohio State University, University of Pittsburgh,
University of Portsmouth, Princeton University, the United States
Naval Observatory, and the University of Washington.

\clearpage

\onecolumn

\onecolumn

\end{document}